\documentclass[12pt,thmsa]{article}
\usepackage{sw20lart}

\input tcilatex
\QQQ{Language}{
American English
}

\begin{document}

\author{Irina Radinschi \thanks{%
radinschi@yahoo.com} and Brindusa Ciobanu \thanks{%
bciobanu2003@yahoo.com} \and Department of Physics, ''Gh. Asachi'' Technical
University, \and Iasi, 700050, Romania}
\title{Weinberg Energy-Momentum Complex for a Stringy Black Hole Solution}
\date{}
\maketitle

\begin{abstract}
In our paper we compute the energy distribution of a magnetic stringy black
hole solution in the Weinberg prescription. The metric under consideration
describes the dual solution in the string frame that is known as the
magnetic stringy black hole solution. The metric is obtained by multiplying
the electric metric in the Einstein frame by a factor $e^{-2\,\Phi }$. The
energy distribution depends on the mass $M$ and charge $Q$. Also, we make a
discussion of the results and we compare our result with those obtained in
the Einstein and Landau and Lifshitz prescriptions and investigate the
connections between the expressions of the energy obtained in these
prescriptions.

Keywords: Weinberg energy-momentum complex, magnetic stringy black hole

PACS: 04. 20 Dw, 04. 70. Bw
\end{abstract}

\section{INTRODUCTION}

The important issue of energy localization still lacks of an acceptable
answer and continues to be one of the most interesting and challenging
problem of the General Relativity. We point out that this problem of
defining in an acceptable manner the energy-momentum density hasn't got a
generally accepted answer yet. The center of general relativity is one of
the most interesting and challenging ideas in modern science, the one that
gravity is the geometry of curved four-dimensional space-time. Mass is the
source of the space-time curvature, and since General Relativity
incorporates Special Relativity, any form of energy is also a source of
space-time curvature. Also, the subject of the localization of energy
continues to be an open one since Einstein [1] has given his important
result of the special theory of relativity that mass is equivalent to energy.

Over the last two decades the issue of the energy-momentum localization by
using the energy-momentum complexes was re-opened and many interesting
results were obtained. In 1990 Bondi [2] gave his opinion that ''a
nonlocalizable form of energy is not admissible in general relativity,
because any form of energy contributes to gravitation and so its location
can in principle be found''. Misner et al [3] sustained that to look for a
local energy-momentum means that is looking for the right answer to the
wrong question. Also, they concluded that the energy is localizable only for
spherical systems. On the other hand, Cooperstock and Sarracino [4]
demonstrated that if the energy is localizable in spherical systems then it
is also localizable in any space-times. The localization of energy is
connected to the use of various energy-momentum complexes, including those
of Einstein [5], Landau and Lifshitz [6], Papapetrou [7], Bergmann [8] and
Weinberg [9]. All of these prescriptions have a drawback: the calculations
are restricted to quasi-Cartesian coordinates. On the other hand, M\o ller
[10] constructed an expression which enables one to calculate the energy
distribution in any coordinate system not only in quasi-Cartesian
coordinates. Also, in this context, of great importance is the Cooperstock
[11] hypothesis which states that the energy and momentum are confined to
the regions of non-vanishing energy-momentum tensor of the matter and all
non-gravitational fields.

As we emphasized above, the problem of energy-momentum localization by using
the energy-momentum complexes was revived at the begining of the last
decade. The interesting results recently obtained by many researchers point
out that the energy-momentum complexes are powerful tools for evaluating the
energy and momentum in a given space-time [12]. Recently, important works
were done with the energy-momentum complexes in 2- and 3-dimensional
space-times [13]. All these considerations demonstrate the significance of
these prescriptions and stress the usefulness of energy-momentum complexes
for the localization of energy.

In our paper we compute the energy distribution of a magnetic stringy black
hole solution in the Weinberg prescription. The metric under consideration
describes the dual solution in the string frame that is known as the
magnetic stringy black hole solution. The metric is obtained by multiplying
the electric metric in the Einstein frame by a factor $e^{-2\,\Phi }$. Also,
we make a discussion of the results and we compare our result with those
obtained in the Einstein and Landau and Lifshitz prescriptions and
investigate the connections between the expressions of the energy obtained
in these prescriptions. Through the paper we use geometrized units ($G=1,c=1$%
) and follow the convention that Latin indices run from $0$ to $3$.

\section{ENERGY\ IN\ THE\ WEINBERG\ PRESCRIPTION}

The study of the stringy black holes is an interesting issue because string
theory is expected to provide us with a finite and clearly defined theory of
quantum gravity and the answer of many questions related with black hole
evaporation could be solved in the context of string theory. String theory
may be the best way to attain the holy grail of fundamental physics, which
is to generate all matter and forces of nature from one basic building
block. Through the years, many important studies have been made related to
the string theory. About the low energy effective theory we point out that
largely resembles general relativity with some new ''matter'' fields as the
dilaton, axion etc [14]-[15]. One of its main property is that there are two
different frames in which the features of the space-time may look very
different. These two frames are the Einstein frame and the string frame and
they are related to each other by a conformal transformation ($g_{\mu \nu
}^E=e^{-2\,\Phi }g_{\mu \nu }^S$) which involves the massless dilaton field
as the conformal factor. The string ''sees'' the string metric. Many of the
important symmetries of string theory also rely of the string frame or the
Einstein frame [16]. The $T$ duality transformation relates metrics in the
string frame only, whereas $S$ duality is a valid symmetry only if the
equations are written in the Einstein frame. Kar [14] gave important results
about the stringy black holes and energy conditions. He studied several
black holes in two and four dimensions with regard to the Weak Energy
Conditions (WEC).

The action for the Einstein-dilaton-Maxwell theory is given by

\begin{equation}
S_{EDM}=\int d^4x\sqrt{-g}e^{-2\,\Phi }[R+4\,g_{\mu \nu }\,\nabla ^\mu
\,\Phi \,\nabla ^\nu \Phi -\frac 12g^{\mu \lambda }g^{\nu \rho }F_{\mu \nu
}F_{\lambda \rho }].  \label{1}
\end{equation}

Varying with respect to the metric, dilaton and Maxwell fields we get the
field equations for the theory given as

\begin{equation}
R_{\mu \nu }=-2\,\nabla _\mu \,\Phi \,\nabla _\nu \Phi +2\,F_{\mu \lambda
}F_\nu ^{\,\,\,\,\lambda }\,,  \label{2}
\end{equation}

\begin{equation}
\nabla ^\nu (e^{-2\,\Phi }F_{\mu \nu })=0,  \label{3}
\end{equation}

\begin{equation}
4\,\nabla ^2\Phi -4(\nabla \Phi )^2+R-F^2=0.  \label{4}
\end{equation}

The metric (in the string frame) which solve the Einstein-dilaton-Maxwell
field equations to yield the electric black hole is given by

\begin{equation}
ds^2=A(1+\frac{2\,M\,\sinh ^2\alpha }r)^{-2}\,dt^2-\frac
1A\,dr^2-r^2\,d\theta ^2-r^2\,\sin ^2\theta \,d\varphi ^2.  \label{5}
\end{equation}

where $A=1-\frac{2\,M}r$.

In the string frame the dual solution known as the magnetic black hole is
obtained by multiplying the electric metric in the Einstein frame by a
factor $e^{-2\,\Phi }$. Therefore, the magnetic black hole metric is given by

\begin{equation}
ds^2=\frac AB\,dt^2-\frac 1{A\,B}\,dr^2-r^2\,d\theta ^2-r^2\,\sin ^2\theta
\,d\varphi ^2,  \label{6}
\end{equation}
with $B=1-$.$\frac{Q^2}{M\,r}$, where $M$ is the mass and $Q$ is the charge
of the magnetic black hole.

The Weinberg energy-momentum complex [9] is given by

\begin{equation}
W^{i\,k}=\frac 1{16\,\pi }D_{\;\,\;\;\,,l}^{l\,i\,k},  \label{7}
\end{equation}

where

\begin{equation}
D_{\;\,\;}^{l\,i\,k}=\frac{\partial h_a^a}{\partial x_l}\eta ^{i\,k}-\frac{%
\partial h_a^a}{\partial x_i}\eta ^{l\,k}-\frac{\partial h^{a\,l}}{\partial
x^a}\eta ^{i\,k}+\frac{\partial h^{a\,i}}{\partial x^a}\eta ^{l\,k}+\frac{%
\partial h^{l\,k}}{\partial x_i}-\frac{\partial h^{i\,k}}{\partial x_l},
\label{8}
\end{equation}

with

\begin{equation}
h_{i\,k}=g_{i\,k}-\eta _{i\,k}  \label{9}
\end{equation}

and $W^{0\,0}$ and $W^{\alpha \,0}$ are the energy and, respectively, the
momentum density components.

The Weinberg energy-momentum complex satisfies the local conservation laws

\begin{equation}
\frac{\partial W^{i\,k}}{\partial x^k}=0.  \label{10}
\end{equation}

Integrating $W^{i\,k}$ over the three-space gives the energy and momentum
components

\begin{equation}
P^i=\int \hskip-7pt\int \hskip-7pt\int W^{i\,0}\,dx^1\,dx^2\,dx^3.
\label{11}
\end{equation}

$P^0$ is the energy and $P^\alpha $ are the momentum components.

Using the Gauss theorem we obtain

\begin{equation}
P^i={\frac 1{16\,\pi }}\int \hskip-7pt\int D_{\;\,\;}^{\alpha
\,0\,i}\,\,n_\alpha \,\,dS,  \label{12}
\end{equation}
where $n_\alpha =\left( {\frac xr},{\frac yr},{\frac zr}\right) $ are the
components of a normal vector over an infinitesimal surface element $%
dS=r^2\sin \theta \,d\theta \,d\varphi $.

For making the calculations using the Weinberg energy-momentum complex we
transform the metric given by (6) to quasi-Cartesian coordinates $t,x,y,z$
according to $x=r\sin \theta \cos \varphi $, $y=r\sin \theta \sin \varphi $
and $z=r\cos \theta $.

Using (7), (8) and applying the Gauss theorem we obtain that the energy
distribution of the magnetic black hole is given by

\begin{equation}
E_W=\frac r2\dfrac{r\,Q^2+2\,M^2\,r-2\,M\,Q^2}{(r-2\,M)(M\,r-Q^2)}.
\label{13}
\end{equation}

The energy distribution depends on the mass $M$ and charge $Q$ of the
magnetic black hole.

We make a comparison with the values of the energy distribution obtained in
the Einstein and Landau and Lifshitz prescriptions. The energy in the
Einstein prescription [17] is given by

\begin{equation}
E_E=\frac 12\dfrac{r\,Q^2+2\,M^2\,r-2\,M\,Q^2}{(M\,r-Q^2)}.  \label{14}
\end{equation}

We establish the connection between the expressions of the energy
distribution in the Weinberg and Einstein prescriptions that is

\begin{equation}
E_W=\frac r{(r-2\,M)}E_E.  \label{15}
\end{equation}

The expression of energy in the Landau and Lifshitz prescription is the same
as in the Weinberg prescription. We conclude that between the Weinberg,
Einstein and Landau and Lifshitz prescriptions there is a relationship given
by

\begin{equation}
E_W=E_{LL}=\frac r{(r-2\,M)}E_E.  \label{16}
\end{equation}

These definitions do not provide the same result for the energy
distribution. It is important that the expression of the energy obtained in
the Weinberg prescription exactly matches with that computed in the context
of the Landau and Lifshitz prescription.

For the Weinberg and Landau and Lifshitz prescriptions we take the limit $%
Q\rightarrow 0$, the Schwarzschild space-time, and we obtain for the energy
distribution the expression

\begin{equation}
E_W=E_{LL}=\dfrac{M^{}}{(1-\frac{2\,M}r)}=M(1-\frac{2\,M}r)^{-1}.  \label{17}
\end{equation}

This expression is the same as obtained by Virbhadra [18] in the case of the
Schwarzschild black hole when the calculations were done in Shwarzschild
Cartesian coordinates.

In the limit $r\rightarrow \infty $ we obtain for the energy distribution of
the magnetic black hole

\begin{equation}
\QATOP{E_W}{r\rightarrow \infty }=\QATOP{E_{LL}}{r\rightarrow \infty }=M+%
\frac{Q^2}{2\,M}.  \label{18}
\end{equation}

\section{DISCUSSION}

The subject of the localization of energy continues to be one of the most
interesting and challenging problems of the General Relativity. This issue
is an open one since Einstein has given his important result of the special
theory of relativity that mass is equivalent to energy. Also, Chang, Nester
and Chen [19] showed that the energy-momentum complexes are actually
quasilocal and legitimate expression for the energy-momentum. They concluded
that there exist a direct relationship between energy-momentum complexes and
quasilocal expressions because every energy-momentum complexes is associated
with a legitimate Hamiltonian boundary term. The concept of energy-momentum
complexes was reconsider by Virbhadra and his collaborators, and since then,
many interesting results was obtained in this area [12]-[13]. Although, the
energy-momentum complexes of Einstein, Landau and Lifshitz, Papapetrou,
Bergmann and Weinberg are coordinate dependent they can give a reasonable
result if calculations are carried out in quasi-Cartesian coordinates.

In this paper we evaluate the energy distribution of a magnetic stringy
black hole solution in the Weinberg prescription. The energy distribution
depends on the mass $M$ and charge $Q$ of the magnetic black hole. We
compare this result with those obtained in the Einstein and Landau and
Lifshitz prescriptions. These definitions do not provide the same result for
the energy of the magnetic stringy black hole, particularly we point out
that the Weinberg and Landau and Lifshitz prescriptions furnish the same
expression for the energy distribution. The connection between the
expressions of the energy distribution obtained in these three prescription
is given by the relation $E_W=E_{LL}=\frac r{(r-2\,M)}E_E$.

These results sustain that the energy-momentum complexes are powerful tools
for obtaining the energy distribution in a given space-time.

\end{document}